\documentclass[aps,superscriptaddress,reprint,10pt]{revtex4-2}

\usepackage{natbib}
\usepackage{url}
\usepackage{textcase}
\usepackage{bm}
\usepackage{times}
\usepackage[top=25truemm,bottom=20truemm,left=20truemm,right=20truemm]{geometry}
\usepackage{mathcomp}
\usepackage{listings}
\usepackage{amsmath}
\usepackage{amsmath,amssymb}
\usepackage{array,booktabs}
\usepackage{ascmac}
\usepackage{bm}
\usepackage{braket}
\usepackage{amsmath}
\usepackage{amsfonts}
\usepackage{cases}
\usepackage{color}
\usepackage{comment}
\usepackage{fancybox}
\usepackage{float}
\usepackage{framed}
\usepackage{graphicx}
\usepackage{here}
\usepackage{hyperref}
\usepackage{mathtools}
\usepackage[version=3]{mhchem}
\usepackage{pifont}
\usepackage{wrapfig}
\usepackage{listings} 
\usepackage{appendix}

\hypersetup{%
 setpagesize=false,%
 bookmarks=true,%
 bookmarksdepth=tocdepth,%
 bookmarksnumbered=true,%
 colorlinks=true,%
 linkcolor=blue,%
 citecolor=blue,%
 urlcolor=blue,%
 pdftitle={},%
 pdfsubject={},%
 pdfauthor={},%
 pdfkeywords={}}

\makeatletter
\let\sectionorig\section
\def\@sectionorig#1{\sectionorig*{\MakeUppercase{#1}}}
\def\@@sectionorig#1{\sectionorig{\MakeUppercase{#1}}}
\renewcommand{\section}{\@ifstar{\@sectionorig}{\@@sectionorig}}
\makeatother

\begin{document} 
\title{A New Scaling Function for QAOA Tensor Network Simulations }
\author{Goro Miki}
\thanks{miki-goro@g.ecc.u-tokyo.ac.jp,\\ Present affiliation: Department of Physics, The University of Tokyo, Tokyo 113–0033, Japan}
\affiliation{Faculty of Pure and Applied Sciences, University of Tsukuba, Tsukuba, Ibaraki 305-8571, Japan}
\affiliation{Global R$\&$D Center for Business by Quantum-AI Technology, 
National Institute of Advanced Industrial Science and Technology (AIST), 1-1-1 Umezono, Tsukuba, Ibaraki 305-8568, Japan}
\author{Yasuhiro Tokura}
\affiliation{Faculty of Pure and Applied Sciences, University of Tsukuba, Tsukuba, Ibaraki 305-8571, Japan}

\begin{abstract}
With the rapid development of quantum computers in recent years, the importance of performance evaluation in quantum algorithms has been increasing. One method that has gained attention for performing this evaluation on classical computers is tensor networks. Tensor networks not only reduce the computational cost required for simulations by using approximations but are also deeply connected to entanglement. Entanglement is one of the most important elements for the quantum advantages of quantum algorithms, but the direct relationship between quantum advantages and entanglement remains largely unexplored. Tensor networks are promising as a means to address this question. In this study, we focus on the entanglement in the Quantum Approximate Optimization Algorithm (QAOA). This study aims to investigate entanglement in QAOA by examining the relationship between the approximation rates of tensor networks and the performance of QAOA. Specifically, we actually perform tensor network simulations of QAOA on a classical computer and extend the study of the scaling relations presented in previous research. We have discovered that scaling relations hold even when entanglement entropy is used as the vertical axis. Furthermore, by analyzing the results of the numerical calculations, we propose a new function for the scaling relation. Additionally, we discovered interesting relationships regarding the behavior of entanglement in QAOA during our analysis. This research is expected to provide new insights into the theoretical foundation of the scaling relations presented in previous studies.
\end{abstract}

\maketitle

\section{Introduction}
In recent years, the rapid development of quantum computers \cite{Feynman, QC1, QC2} has led to active research on hardware aimed at practical implementation. A quantum computer is a computational device that utilizes the principles of quantum mechanics, specifically entanglement and superposition, to perform calculations. It is expected that for certain problems, quantum computers can execute computations significantly faster than classical computers \cite{QC3}. Currently, quantum computers with hundreds of qubits have already been realized \cite{IBMQ}, and the number of qubits is expected to continue increasing in the future. However, a quantum computer that surpasses classical computers, including supercomputers, in solving practical problems such as prime factorization has yet to be realized. The primary reason for this is that solving complex problems like prime factorization on a quantum computer necessitates the implementation of quantum error correction \cite{Error correction, EC2}.

In quantum error correction, computations are performed using logical qubits,
which are constructed from groups of physical qubits capable of correcting errors. At present, implementing quantum error correction on quantum hardware remains an extremely challenging technology \cite{EC3}. Furthermore, quantum error correction requires additional qubits, which increases the number of qubits needed for computation, posing another obstacle to its realization. The quantum computers that currently exist are those that do not incorporate quantum error correction, commonly referred to as Noisy Intermediate-Scale Quantum (NISQ) computers \cite{NISQ}.

It is known that almost all quantum algorithms \cite{Grover, NC}, including Shor's algorithm \cite{Shor}, a quantum algorithm for solving prime factorization, cannot be executed accurately on current NISQ computers. Given this background, there is an increasing demand for research investigating the error rates required for quantum hardware to enable the success of quantum algorithms. The most effective approach for such research is to simulate quantum circuits on classical computers. By simulating quantum algorithms on classical computers and evaluating their performance, researchers can gain insights into the characteristics of quantum algorithms.

Among the most promising methods for enabling classical simulation of quantum algorithms is the tensor network approach \cite{TN, TN2, TN3}. Originally developed for simulating quantum many-body systems on classical computers, tensor networks remain an actively researched area. The key advantage of tensor network simulations is their ability to efficiently approximate quantum states. One of the main challenges in simulating quantum states on computers is that an enormous number of parameters are required to represent large systems. This makes classical simulation of large quantum systems difficult. Tensor networks overcome this challenge by approximating quantum states, reducing the number of parameters needed for simulation. Singular value decomposition (SVD) \cite{SVD} is used to decompose quantum states into tensor products, and by selectively retaining only the most significant singular values, an efficient approximation of quantum states is achieved.

While tensor networks were originally used primarily for simulating spin systems, the growing interest in quantum computing has led to their increased use in quantum circuit simulation. One of the reasons tensor networks have garnered attention is their deep connection to entanglement in quantum circuits \cite{TN}. Entanglement is a crucial factor in accelerating quantum algorithms, and many quantum algorithms leverage its properties. However, the exact role of entanglement in enhancing quantum computational speed remains an open question. Since tensor network simulations explicitly reveal the amount of entanglement in a system, they are considered a suitable tool for investigating the role of entanglement in quantum algorithms. Understanding how entanglement contributes to the speed and performance of quantum algorithms is essential for advancing our knowledge of quantum computation and accelerating the realization of practical quantum computers.

The Quantum Approximate Optimization Algorithm (QAOA) \cite{QAOA1, QAOA2} is a specialized algorithm for solving combinatorial optimization problems, such as the traveling salesman problem. Since it is executed using both quantum and classical computers, it is also referred to as a quantum-classical hybrid algorithm. 
In quantum mechanical terms, QAOA is used to determine the ground state and its energy of a given Hamiltonian. A notable feature of QAOA is that it uses relatively shallow quantum circuits compared to other quantum algorithms. A shallow quantum circuit contains fewer quantum gates, which is desirable in quantum hardware implementation because longer circuits accumulate more errors, making hardware realization more challenging. However, from the perspective of quantum advantage (i.e., the ability to outperform classical algorithms), shallow circuits present a drawback: they increase the likelihood that the algorithm can be efficiently simulated on a classical computer, thereby diminishing the need for quantum hardware. Given this background, there are many doubts about whether QAOA truly possesses quantum advantage.

A previous study \cite{Scaling1} demonstrated, within the framework of tensor networks, that QAOA's quantum advantage can be quantitatively evaluated. Specifically, the authors of this study showed that the relationship between the approximation ratio of tensor networks and the performance of QAOA can be expressed as a single curve independent of the number of qubits. They referred to this relationship as a scaling relation. In other words, this study clearly identified the threshold at which classical computers can effectively simulate QAOA, making it a groundbreaking discovery.
In deriving the scaling relation, the authors employed Matrix Product States (MPS) \cite{MPS}, the simplest tensor network method, to perform classical simulations of QAOA. The study specifically examined how the energy expectation values obtained from QAOA depend on the approximation ratio of the tensor network. The results aligned with intuition: as the approximation ratio decreased, the energy expectation value deviated further from the correct ground energy. However, the nontrivial finding of this study was that the variation in energy expectation values could be expressed as a universal function independent of the number of qubits. The authors of the previous study identified a clear theoretical explanation for this result as an open question.

In this paper, we aim to elucidate the theoretical foundation of the scaling relation proposed in a previous study. Understanding this theoretical basis will clarify the fundamental characteristics of QAOA as a quantum algorithm, which, in turn, is expected to lead to further discoveries regarding its quantum advantage.
In this study, we extended the scaling relation proposed in previous research \cite{Scaling1} and discovered a new scaling relation based on entanglement entropy. Since this scaling relation is based on entanglement entropy, a well-defined quantity, it allows for theoretical analysis. We derived the scaling function by analyzing our results. To the best of our knowledge, this is the first theoretical analysis of the scaling relation in QAOA. Furthermore, numerical results used to derive the scaling function show that the entanglement entropy in QAOA initially increases linearly and then stabilizes. This behavior is likely specific to optimized QAOA circuits. However, if other quantum algorithms exhibit similar entanglement entropy behavior, analogous scaling relations may also apply.

\section{Background}
\subsection{Motivation for Using MPS}
 In this section, we briefly review the motivation for using tensor networks, particularly Matrix Product States (MPS) \cite{MPS}. First, a general quantum state can be described as follows:
\begin{equation}
\label{1}
    \ket{\Psi}=\sum_{i_1,\cdots, i_N}\Psi_{i_1 i_2 i_3\cdots i_N}\ket{i_1 i_2 i_3\cdots i_N}
\end{equation}
 Here, $N$ represents the size of the system, which corresponds to the number of qubits in the case of
 quantum computation. In the case of qubits, $i_1, i_2, i_3, ..., i_N$ all have a degree of freedom of 2
 (taking values of either 0 or 1). The number of parameters required to describe this quantum state is
 at most $2^N$. This means that the number of parameters required to represent a quantum state
 increases exponentially with $N$, making it difficult to simulate large $N$ cases on classical computers.
 On the other hand, by focusing on the coefficient $\Psi_{i_1 i_2 ... i_N}$, it is known that this coefficient
 can be decomposed into a product of tensors as follows:
\begin{equation}
\label{2}
    \Psi_{i_1 i_2 i_3\cdots i_N}=\sum_{\alpha_0,\cdots, \alpha_N}A^{i_1}_{\alpha_0\alpha_1}A^{i_2}_{\alpha_1\alpha_2}A^{i_3}_{\alpha_2\alpha_3}\cdots A^{i_N}_{\alpha_{N-1}\alpha_N}
\end{equation}
 This expression is a product of $N$ tensors $A^i_{\alpha_{n-1} \alpha_n}$ ($n = 1,\cdots, N$). The summation
 over variables $\alpha_0, \cdots, \alpha_N$ is called contraction, which corresponds to the connection
 between the tensors.

By substituting the coefficients Eq. (\ref{2}) into the quantum state Eq. (\ref{1}), we obtain the new form of the quantum state.
 This representation is called the Matrix Product State (MPS) \cite{MPS}. MPS is the simplest form of tensor
 networks, but its fundamental concepts apply to other tensor networks as well. Here, $\alpha_0$ and
 $\alpha_N$ have a degree of freedom of 1, so they can be omitted. Also, let the degree of freedom of
 $\alpha_n$ ($n: 1,\cdots, N-1$) be $\chi_n$ (i.e., $\alpha_n: 1,\cdots, \chi_n$). Each $\chi_n$ is called the bond dimension.
 To analyze the number of parameters, we assume that each $\alpha_n$ has different degrees of
 freedom and define $\chi$ as the maximum, i.e., $\chi = \text{max}(\chi_n$). The maximum number of
 parameters required to represent this quantum state is $2N\chi^2$.
 Comparing this to the original representation requiring $2^N$ parameters, we see that MPS reduces
 the parameter count from an exponential to a polynomial scale. However, if $\chi$ grows exponentially
 with $N$, the reduction effect is nullified. For quantum states where $\chi$ does not grow exponentially,
 MPS is highly effective.
 The scaling of $\chi$ with respect to $N$ is related to the amount of entanglement in the quantum state.
 The greater the entanglement, the larger the $\chi$ value. In a quantum state with no entanglement, $\chi
 = 1$.

Decomposing a tensor product as in Eq. (\ref{2}) requires a method called singular value decomposition (SVD) \cite{SVD}. 
SVD is a generalized form of diagonalization, which allows any matrix to be decomposed into the product of two unitary matrices and one diagonal matrix. 
The singular value decomposition of an \( m \times n \) matrix \( M \) is expressed as follows:
\begin{equation}
\label{3}
M = U \Sigma V^\dagger
\end{equation}
Here, if we define \( k = \min(m, n) \), then \( U \) is an \( m \times k \) unitary matrix, \( V \) is an \( n \times k \) unitary matrix, and \( V^\dagger \) is the Hermitian conjugate of \( V \). 
The middle matrix \( \Sigma \) is a \( k \times k \) square diagonal matrix. 
The diagonal elements are arranged in descending order of their singular values, which are always non-negative. 
If we denote the number of nonzero singular values as \( r \), and the singular values as \( \lambda_1, \lambda_2, \dots, \lambda_r \), then \( \Sigma \) takes the following form:
\begin{equation}
\label{4}
\Sigma =
\begin{bmatrix}
\lambda_1 & 0 & \dots & 0 & \dots & 0 \\
0 & \lambda_2 & \dots & 0 & \dots & 0 \\
\vdots & \vdots & \ddots & \vdots & \dots & \vdots \\
0 & 0 & \dots & \lambda_r & \dots & 0 \\
\vdots & \vdots & \dots & \vdots & \ddots & \vdots \\
0 & 0 & \dots & 0 & \dots & 0
\end{bmatrix}
\end{equation}
Previously, we mentioned the relationship between entanglement and bond dimension. It is also known that there is a connection between entanglement and the distribution of singular values $\{\lambda_1,\lambda_2,\cdots,\lambda_r,0,\cdots,0\}$. In summary, the greater the entanglement, the more uniformly distributed all singular values become. When entanglement is maximized, all singular values are equal. That is, in the matrix of Eq. (\ref{4}), there are no zero singular values (i.e., $r=k$), and $\lambda_1=\lambda_2=\cdots=\lambda_r$. 

Conversely, as the amount of entanglement decreases, the proportion occupied by larger singular values (such as $\lambda_1,\lambda_2$) increases relative to the entire set, while the smaller singular values $\lambda_{r-1},\lambda_r$ become nearly zero. In this case, it is known that assuming the lower singular values are strictly zero does not significantly reduce the information contained in the quantum state. In other words, the original quantum state can be approximated with high accuracy. This process is generally referred to as low-rank approximation or truncation. Truncation is one of the most important operations in tensor networks and is widely used not only in matrix product states but also in other tensor network methods. By performing this operation, a smaller matrix—with fewer parameters—can approximate the quantum state, allowing classical computers with limited memory capacity to represent complex quantum states and simulate their dynamics.

\subsection{The relationship between MPS and entanglement}
Now, we will review the relationship between MPS and entanglement. 
The bond dimensions $\chi_n\ (n:1,\cdots,N-1)$ in Eq. (\ref{2}) correspond to the amount of information that can be stored in the MPS and indicate the amount of entanglement in the MPS.
In the following discussion, we mathematically confirm this point. First, we consider a Hilbert space that is divided into two parts as $H=H_L \otimes H_R$, with the boundary between the $n$-th tensor and the $(n+1)$-th tensor. Here, $H_L (H_R)$ represents the subspace on the left (right) side of the boundary between the $n$-th tensor and the $(n+1)$-th tensor.
Decomposing the state $\ket{\Psi}$ across the boundary between the $n$-th and $(n+1)$-th tensors, we obtain the following expression:
\begin{equation}
\label{Schmidt}
    \ket{\Psi}=\sum^{d}_{k=1}\lambda_k\ket{\Psi^L_k}\ket{\Psi^R_k}
\end{equation}
Here, $\ket{\Psi^{L(R)}_k} \in H_{L(R)}$. The decomposition in the form of Eq. (\ref{Schmidt}) is called the Schmidt decomposition. The parameter $d$ represents a quantity related to the bond dimension of the partitioned boundary and is expressed mathematically as $d=\text{min}(\chi_n, 2\chi_{n+1})$.
The values $\lambda_k \ (k:1,\cdots,d)$ are the singular values obtained when applying singular value decomposition to the three-leg tensor $A^{i_{n+1}}_{\alpha_n\alpha_{n+1}}$, considering it as a $\chi_n \times 2\chi_{n+1}$ matrix.

Next, using the Schmidt decomposition of Eq. (\ref{Schmidt}), we calculate the entanglement entropy to examine the amount of entanglement between the subspaces $L$ and $R$. The entanglement entropy is the most important quantity in this paper and will frequently appear in subsequent discussions. The definition of the entanglement entropy $S$ is given as follows:
\begin{equation}
\label{S1}
    S=-\text{Tr}_R[\rho_R \log_2(\rho_R)]=-\text{Tr}_L[\rho_L \log_2(\rho_L)]
\end{equation}
Here, $\rho_L$ and $\rho_R$ are the reduced density matrices defined as $\rho_{R(L)}=\text{Tr}_{L(R)}(\rho), \ \rho=\ket{\Psi}\bra{\Psi}$. Using the Schmidt decomposition of $\ket{\Psi}$ in Eq. (\ref{Schmidt}), $S$ is calculated as follows:
\begin{equation}
\label{S2}
    S=-\sum^{d}_{k=1}\lambda^2_k \log_2\lambda^2_k
\end{equation}
By using the method of Lagrange multipliers (the constraint condition is $\text{Tr}_L(\rho_L)=\text{Tr}_R(\rho_R)=1$), it is found that the maximum value of the entanglement entropy is given by \(\lambda_k = 1 / \sqrt{d}\), which takes the maximum value of \(\log_2{(d)}\). 
In general, it is reasonable to assume that \(\chi_n \leq 2 \chi_{n+1}\). 
From this fact, the maximum entanglement entropy between the $n$-th and $(n+1)$-th tensors is expressed as follows:
\begin{equation}
\label{Smax}
    S_{\text{max}}=\log_2{\chi_n}
\end{equation}
Based on the above, it has been confirmed that the maximum value of entanglement entropy depends on the bond dimension, and that the entanglement entropy is maximized when all singular values are equal. The distribution of singular values depends on the amount of entanglement, and it is known that the singular values decrease at a faster rate when the entanglement is smaller.

\subsection{Quantum Approximate Optimization Algorithm}
This subsection reviews the Quantum Approximate Optimization Algorithm (QAOA) \cite{QAOA1, QAOA2}, which plays a
 crucial role in this paper. 
QAOA is a type of quantum-classical hybrid algorithm designed primarily for solving combinatorial optimization problems, such as the traveling salesman problem.

Consider a function $C(\mathbf{z})$ determined by an $N$-bit binary sequence $\mathbf{z} = z_1z_2\cdots z_N$ with $z_i \in \{-1,1\}$. The optimization problem can be reduced to finding the sequence $\mathbf{z}$ that maximizes (or minimizes) a given classical objective function $C(\mathbf{z}): \{+1, -1\}^N \mapsto \mathbb{R}_{\ge 0}$. 
The simplest method to solve this problem is to compute $C(\mathbf{z})$ for all possible $\mathbf{z}$ and determine the maximum (or minimum) value. In this case, since each $z_i$ takes values in $\{-1,1\}$ and there are $N$ such elements, the worst case requires evaluating $C(\mathbf{z})$ up to $2^N$ times. This implies that the computational complexity grows exponentially with $N$, making it impractical to solve for large $N$.
To address this, QAOA aims to solve the optimization problem using a quantum computer. It is believed that a quantum computer can mitigate the exponential computational complexity; however, it has not been mathematically proven that QAOA can solve optimization problems in polynomial time. 

In QAOA, to execute a combinatorial optimization problem using quantum computation, the binary variables $z_i$ in the classical objective function $C(\mathbf{z})$ are transformed into quantum spins $\sigma^z_i$. Through this transformation, $C(\mathbf{z})$ can be regarded as the following Hamiltonian:
\begin{equation}
\label{HC}
    H_C=C(\sigma^1_z, \sigma^2_z, \cdots ,\sigma^N_z)
\end{equation}
In QAOA, in addition to the Hamiltonian in Eq. (\ref{HC}), another Hamiltonian is introduced. This additional Hamiltonian always has the same form, regardless of the problem that QAOA is designed to solve. This Hamiltonian can be written as follows:
\begin{equation}
    H_B=\sum^N_{j=1}\sigma^j_x
\end{equation}
In the following, we construct a time evolution operator using these two Hamiltonians and convert it into a quantum circuit.

First, we prepare an $N$-qubit circuit, where $N$ corresponds to the number of variables $z_i$. The initial state of the quantum circuit is set to $\ket{+}^{\otimes N}$. 
Next, we apply a quantum circuit corresponding to the operator $e^{-i\beta_1 H_B}e^{-i\gamma_1 H_C}$ on this initial state. Here, $\beta_1$ and $\gamma_1$ are arbitrary parameters.
After executing this circuit, another quantum circuit corresponding to the operator $e^{-i\beta_2 H_B}e^{-i\gamma_2 H_C}$ is applied to the resulting quantum state. The parameters $\beta_2$ and $\gamma_2$ are distinct from $\beta_1$ and $\gamma_1$. This process is repeated $p$ times with different parameters. The number of repetitions $p$ is generally referred to as the number of QAOA layers. After $p$ iterations, the quantum state $\ket{\Psi_p(\Vec{\gamma},\Vec{\beta})}$ is given as follows:
\begin{equation}
    \ket{\Psi_p(\Vec{\gamma},\Vec{\beta})}=e^{-i\beta_p H_B}e^{-i\gamma_p H_C}\cdots e^{-i\beta_1 H_B}e^{-i\gamma_1 H_C}\ket{+}^{\otimes N}
\end{equation}
The parameterized quantum circuit used to generate this wavefunction is called the QAOA circuit (or QAOA ansatz).

The expectation value of $H_C$ is calculated as:
\begin{equation}
    F_p(\Vec{\gamma},\Vec{\beta})=\braket{\Psi_p(\Vec{\gamma},\Vec{\beta})|H_C|\Psi_p(\Vec{\gamma},\Vec{\beta})}
\end{equation}
The calculation of the expectation value is feasible on a quantum computer. The expectation value $F_p(\Vec{\gamma},\Vec{\beta})$ that we have obtained also contains $2p$ parameters $\gamma_i, \beta_i \ (i=1,2,\cdots,p)$.
The reason for introducing these parameters is to perform optimization with respect to them. This optimization will be executed on a classical computer. 
The newly updated parameters are denoted as $\gamma^*_i$ and $\beta^*_i$ ($i = 1, 2, \cdots, p$).
Using these updated parameters, a new QAOA circuit is constructed. 
Then, applying this new QAOA circuit to the initial state $\ket{+}^{\otimes N}$, 
a new wavefunction $\ket{\Psi_p(\Vec{\gamma^*},\Vec{\beta^*})}$ is created, 
and the expectation value $F_p(\Vec{\gamma^*},\Vec{\beta^*})$ is calculated.
The parameter updates are then applied again to $F_p(\Vec{\gamma^*},\Vec{\beta^*})$.
This process is repeated until the wavefunction converges to the desired ground state $\ket{\Psi_p(\Vec{\gamma}_{\text{opt}},\Vec{\beta}_{\text{opt}})}$.
This is the basic procedure of QAOA.

\subsection{MAXCUT problem}
In this paper, we focus on solving the MAXCUT problem using QAOA.
In the MAXCUT problem, an arbitrary graph is given. Let $N$ be the number of vertices in the given graph. 
Each vertex is assigned a random number. The graph contains multiple edges connecting the vertices, and each edge is assigned a weight. 
Here, let $\braket{i,j}$ represent all possible edge pairs in the graph (i.e., edges connecting vertex $i$ and vertex $j$). 
The weight of each edge is denoted as $w_{ij}$. 
In the MAXCUT problem, the goal is to divide the vertices into two separate groups by drawing a single cut through the graph. 
The MAXCUT problem is defined as finding the cut that maximizes the sum of the edge weights across the division. 
Since this problem involves finding the maximum cut, it is called the MAXCUT problem.

Next, we discuss how to mathematically formulate the MAXCUT problem. 
This problem is equivalent to minimizing the following function:
\begin{equation}
    C(\mathbf{z})=\sum_{<i,j>}w_{ij}z_i z_j
\end{equation}
where $z_i$ and $z_j$ take values of either $+1$ or $-1$. 
When drawing a cut to separate the vertices into two groups, we define $z_i = +1$ if vertex $i$ belongs to one group and $z_i = -1$ if it belongs to the other group
(this assignment of $+1$ and $-1$ is arbitrary and can be reversed). Under this definition, when vertices $i$ and $j$ belong to different groups, $w_{ij}z_i z_j=-w_{ij}$, and when they belong to the same group, $w_{ij}z_i z_j=w_{ij}$. 
Therefore, finding the minimum of $C(\mathbf{z})$ is equivalent to solving the MAXCUT problem. 
Furthermore, to apply QAOA, the MAXCUT problem needs to be expressed in quantum computational form. 
The MAXCUT problem is equivalent to finding the minimum value of the following Hamiltonian:
\begin{equation}
\label{HC2}
    H_C=\sum_{<i,j>}w_{ij}\sigma^i_z\sigma^j_z
\end{equation}
Finding the ground state of $H_C$ corresponds to finding the solution to the MAXCUT problem, which can be achieved using QAOA.

There are various types of graphs used in solving the MAXCUT problem. 
However, in this study, we focus only on a special type of graph called a complete graph, 
where every pair of vertices is connected by an edge. In this paper, the edge weights $w_{ij}$ are randomly chosen from values between 0 and 1. 

\subsection{Scaling relations in energy expectation values}

This subsection discusses the findings of previous research \cite{Scaling1, Scaling2}. In \cite{Scaling1}, the MAXCUT problem in several types of graphs was solved using QAOA, 
and the QAOA circuit was simulated using MPS. The authors of \cite{Scaling1} reported the following scaling relation by comparing the energy expectation values of QAOA circuits with and without truncation:
\begin{equation}
\label{Scaling1}
    \frac{\braket{\Psi^{\chi}_p(\Vec{\gamma}_{\text{opt}},\Vec{\beta}_{\text{opt}})|H_C|\Psi^{\chi}_p(\Vec{\gamma}_{\text{opt}},\Vec{\beta}_{\text{opt}})}}{\braket{\Psi_p(\Vec{\gamma}_{\text{opt}},\Vec{\beta}_{\text{opt}})|H_C|\Psi_p(\Vec{\gamma}_{\text{opt}},\Vec{\beta}_{\text{opt}})}}=\mathcal{F}\left(\frac{\ln{\chi}}{N}\right)
\end{equation}
Here, $\mathcal{F}$ is an undetermined function, $\chi$ is the number of singular values retained during truncation, 
and $N$ is the number of qubits (or equivalently, the number of vertices in the graph of the MAXCUT problem). 
The denominator on the left-hand side of Eq. (\ref{Scaling1}) is the energy expectation value obtained when QAOA is executed without truncation. Specifically, the output energy expectation value is calculated as follows: after the optimization cycle of QAOA is completed and the result is output, the optimized parameters $(\Vec{\gamma}_{\text{opt}},\Vec{\beta}_{\text{opt}})$ are used to construct the wave function $\ket{\Psi_p(\Vec{\gamma}_{\text{opt}},\Vec{\beta}_{\text{opt}})}$ via the QAOA circuit (which is simulated using MPS), and then the expectation value of the Hamiltonian $H_C$ is computed with this wave function. In the case of the MAXCUT problem, the problem Hamiltonian is given by Eq. (\ref{HC2}). The output energy expectation value is $\braket{\Psi_p(\Vec{\gamma}_{\text{opt}},\Vec{\beta}_{\text{opt}})|H_C|\Psi_p(\Vec{\gamma}_{\text{opt}},\Vec{\beta}_{\text{opt}})}$.

On the other hand, the numerator corresponds to the energy expectation value obtained when truncation is applied, given by $\braket{\Psi^{\chi}_p(\Vec{\gamma}_{\text{opt}},\Vec{\beta}_{\text{opt}})|H_C|\Psi^{\chi}_p(\Vec{\gamma}_{\text{opt}},\Vec{\beta}_{\text{opt}})}$. Here, $\chi$ is the bond dimension when truncation is applied (i.e., only $\chi$ singular values are kept, with all other singular values set to zero).
When truncation is performed, the process follows as: after the optimization cycle of QAOA is completed and the result is output, the QAOA circuit (which is simulated using MPS) is truncated to create an approximate wave function $\ket{\Psi^{\chi}_p(\Vec{\gamma}_{\text{opt}},\Vec{\beta}_{\text{opt}})}$ (the parameters ($\Vec{\gamma}_{\text{opt}},\Vec{\beta}_{\text{opt}}$) are the same as those used when no truncation was applied), and the expectation value of the Hamiltonian $H_C$ is calculated with this wave function. Note that the QAOA circuit in the cycle does not undergo truncation, and truncation is only applied to the final circuit of the cycle. 

Additionally, the paper \cite{Scaling2} demonstrated that the same scaling relation holds for complete graphs with 1 to 6 QAOA layers, 3-regular graphs with 1 or 2 QAOA layers, and 4-regular graphs with 1 or 2 QAOA layers.
In all cases, the function increases monotonically and then saturates.
Furthermore, \cite{Scaling1} reports that the scaling relation also holds when the bipartite von Neumann entanglement entropy from a cut in the middle of the MPS divided by 
$N$ is used as the vertical axis.

The reason this is called a ``scaling relation" is that data from all different $N$ values align on a single
 universal curve. 
In the MAXCUT problem, different values of $N$ correspond to different problem instances. Nevertheless, 
Eq. (\ref{Scaling1}) show that the values align on a single curve. 
However, the theoretical basis for this scaling relation remains unclear, as noted by the authors of \cite{Scaling1,Scaling2}. 
To the best of our knowledge, no answer has been proposed so far. In our study, we extend the findings of 
previous research and explore the theoretical foundation of the scaling relation. The details will be discussed in the next section.

\section{Results}
\normalsize
This study extends the previous research \cite{Scaling1,Scaling2} and adopts a new vertical axis instead of the energy expectation value to explore the theoretical basis of the scaling relations in Eq. (\ref{Scaling1}). We consider only the MAXCUT problem on a complete graph and one QAOA layer in the following.
The adopted vertical axis is defined as the ratio of the sum of the entanglement entropy over all bonds in all MPSs immediately after applying CNOT gates in the QAOA circuit, comparing cases with and without truncation. Hereafter, we denote entanglement entropy as EE for brevity. 
We confirm that the scaling relation holds even when using this vertical axis. Expressing this mathematically, we obtain the following equation ($\lambda_{i,j,k}$ represents the singular values):
\begin{equation}
\begin{split}
\label{Scaling for EE}
    \frac{\sum S^{\chi}}{\sum S}&\equiv \frac{\sum^{N-1}_{i=1} \sum^{N(N-1)}_{j=1} \left(-\sum^{\chi}_{k=1} \lambda^2_{i,j,k}\log_2{\lambda^2_{i,j,k}}\right)}{\sum^{N-1}_{i=1} \sum^{N(N-1)}_{j=1} \left(-\sum^{2^{N/2}}_{k=1} \lambda^2_{i,j,k}\log_2{\lambda^2_{i,j,
,k}}\right)}\\
&=\mathcal{G}\left(\frac{2\log_2{\chi}}{N}\right)
\end{split}
\end{equation}
$\Sigma_i$ represents the summation over all bonds in the MPS, and $\Sigma_j$ represents the summation over all time steps in the QAOA circuit. 
The reason for adopting this vertical axis is that, as shown in Appendix \ref{Ap1}, examining EE at just a single point (a single bond) does not satisfy the scaling relation.

\begin{figure}[h]
  \centering\vspace{(0mm)}
  \includegraphics[width=0.9\linewidth]{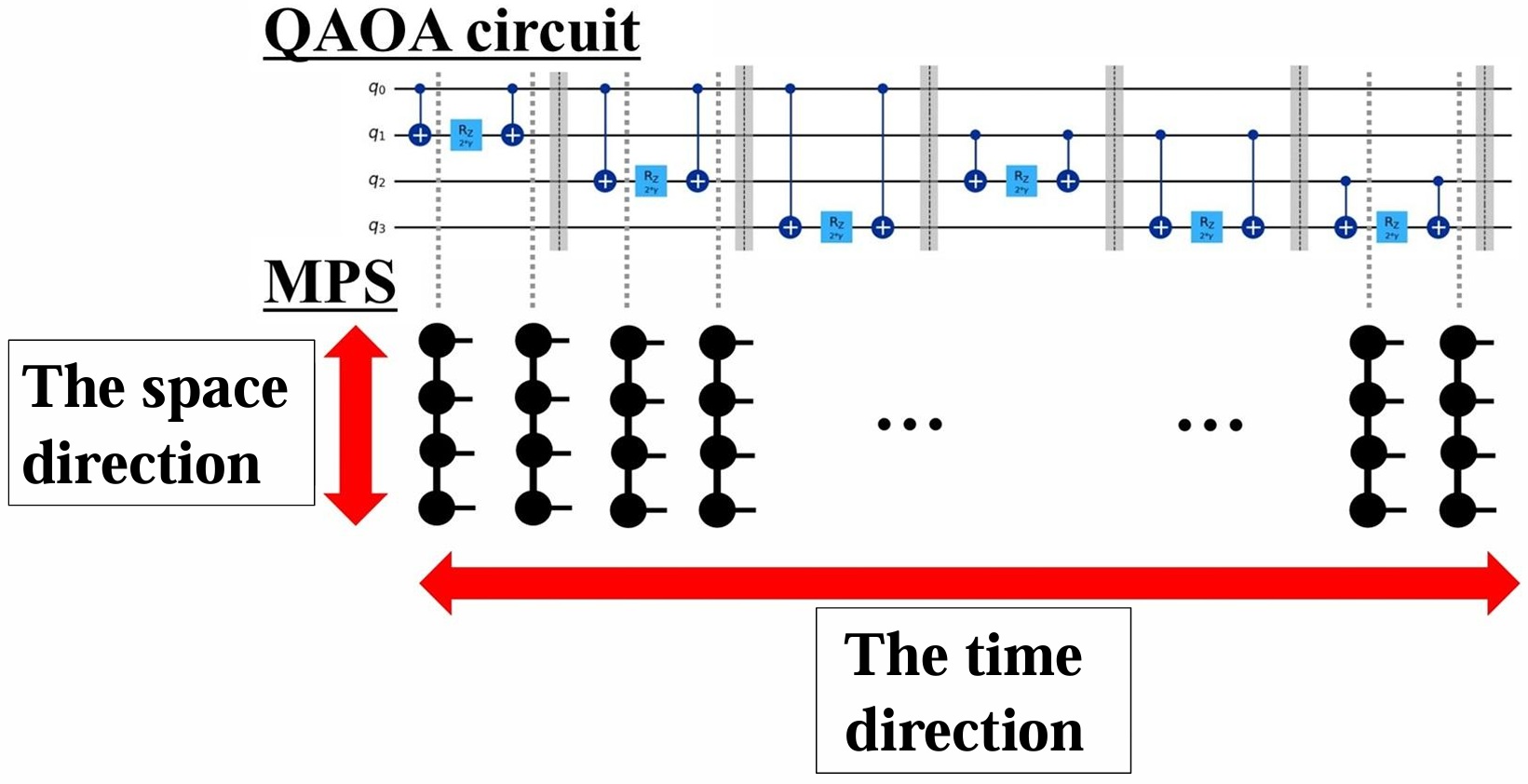}
  \caption{An illustrative diagram of the `space direction' and `time direction' of the MPS corresponding to the QAOA circuit.}
  \label{fig:Space and time}
\end{figure}

\begin{figure}[h]
  \centering\vspace{(0mm)}
  \includegraphics[width=0.85\linewidth]{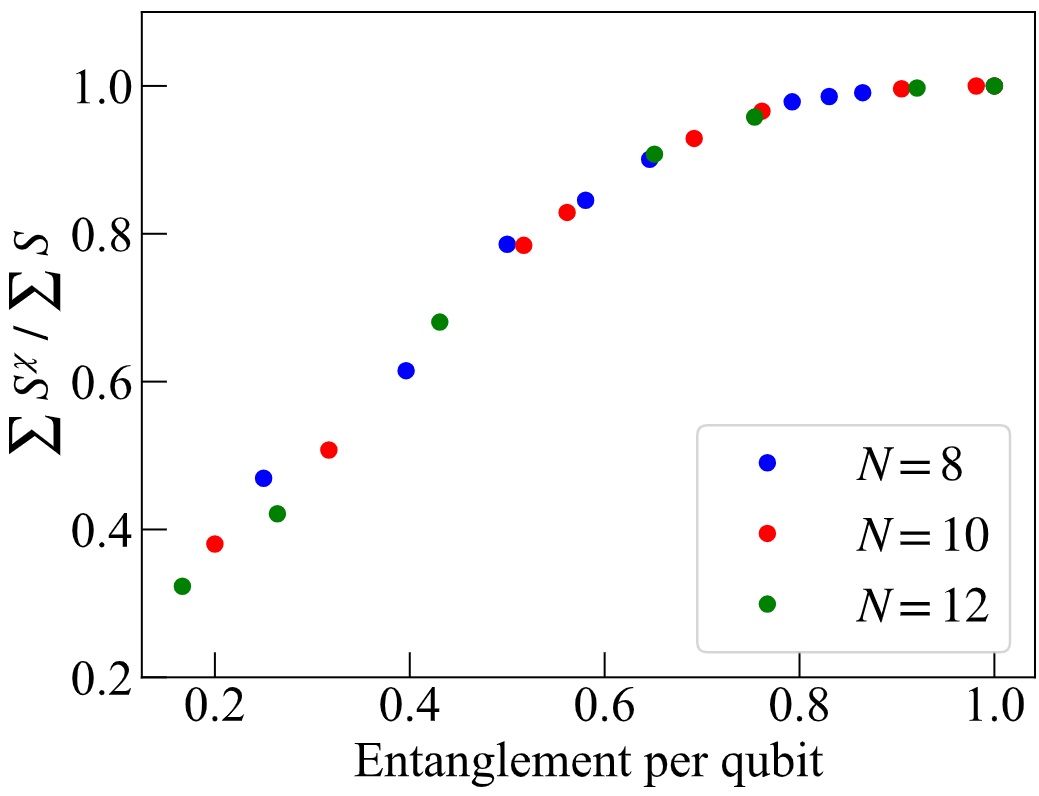}
  \caption{The scaling relation in the case of Eq. (\ref{Scaling for EE}) (an example with a complete graph and one QAOA layer). The average is taken over 100 samples with respect to the randomness of edge weights in the graph ($w_{i,j}\in [0,\ 1]$ in Eq. (\ref{HC2})).}
  \label{fig:Scaling Fig1}
\end{figure}

The summation of EE over the bonds is illustrated in Fig. \ref{fig:Space and time}, where EE is summed over both the `space direction' and `time direction'.
We explain this EE summation using the example of $N=8$.
First, we consider the number of CNOT gates present in the QAOA circuit for a complete graph. The number of CNOT gates is twice the number of edges in the graph corresponding to the MAXCUT problem. The number of edges in a complete graph is given by $\frac{N(N-1)}{2}$. Therefore, for $N=8$, the number of CNOT gates in the circuit is 
$56$.
The 56 MPSs immediately after applying these CNOT gates correspond to the `time direction' in Fig. \ref{fig:Space and time}. Since each MPS contains $7$ bonds, we calculate EE for all these bonds, corresponding to the `space direction' in the figure. 
Thus, for a complete graph with $N=8$, we compute EE for $56\cdot 7 =392$ bonds and take the total sum. The same summation is performed in the truncated case. Then, the ratio is calculated as in Eq. (\ref{Scaling for EE}).
The scaling relation in Eq. (\ref{Scaling for EE}) is plotted in Fig. \ref{fig:Scaling Fig1}. The problem considered here is the example of QAOA with one layer for the MAXCUT problem on a complete graph. The details of the simulation are described in Appendix \ref{Simulation1}.

\section{Analysis}
\normalsize
In this section, we perform a theoretical analysis of the obtained results, specifically Eq. (\ref{Scaling for EE}) and Fig. \ref{fig:Scaling Fig1}.
First, we calculate the theoretical maximum of the summation of EE over all bonds in the MPS immediately after applying a CNOT gate at a certain time step in the QAOA circuit. Hereafter, we refer to this summation of EE as the ``summation of EE in the space direction" and denote it as $\sum_{\text{space}} S$.

It is known that the summation of EE in the space direction reaches its maximum when the bond dimensions increase as $2,4,8,\dots,2^{N/2}$ from the edges toward the center in the MPS (In the following, $N$ is assumed to be even). This exponential increase in bond dimensions follows from the properties of SVD. 
Furthermore, when the bond dimension at a certain bond is $\chi_n$, the maximum value of its entanglement entropy is given by $\log_2{\chi_n}$, as derived in Eq. (\ref{Smax}) (this corresponds to the case where all singular values are equal). From this, the theoretical maximum of the summation of EE in the space direction, $\left(\sum_{\text{space}} S \right)_{\text{max}}$, can be calculated as follows:
\begin{equation}
\begin{split}
\label{Smax theory}
    \left(\sum_{\text{space}}S \right)_{\text{max}}&=2\sum_{k=1}^{\frac{N}{2}-1} \log_2{2^k}+\frac{N}{2}\\
    &=\frac{N^2}{4}
\end{split}
\end{equation}

\begin{figure}[h]
  \centering\vspace{(0mm)}
  \includegraphics[width=0.85\linewidth]{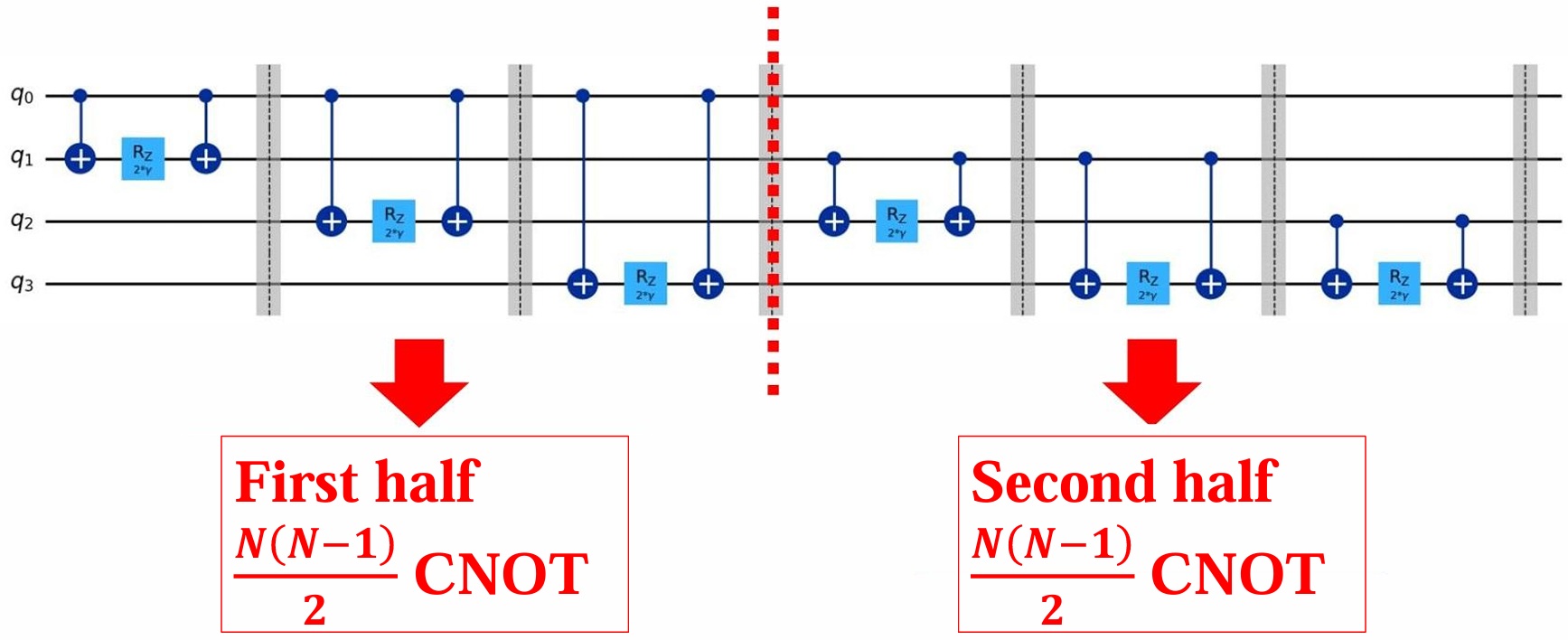}
  \caption{An illustrative diagram of how to divide the QAOA circuit into the first and second halves in the case without truncation.}
  \label{fig:first and second 1}
\end{figure}
 Here, we introduce a certain condition based on numerical calculations.
First, as shown in Fig. \ref{fig:first and second 1}, we consider splitting the number of CNOT gates in the circuit into two halves. It was previously stated that when the number of QAOA layers is 1, the number of CNOT gates in the QAOA circuit for solving the MAXCUT problem on a complete graph is $N(N-1)$. That is, we define the number of CNOT gates in the first half of the circuit as $\frac{N(N-1)}{2}$ and the number of CNOT gates in the second half as $\frac{N(N-1)}{2}$.
Next, we compute the average of $\sum_{\text{space}} S$ in both the first and second halves of the circuit and denote them as $(\overline{\sum_{\text{space}} S})_{\text{first}}$ and $(\overline{\sum_{\text{space}} S})_{\text{second}}$, respectively. 
From numerical calculations, the results shown in Table \ref{table1} were obtained for these two values.

\Large
\begin{table}[h]
    \centering
    \caption{Verification in numerical calculations. The average is taken over 100 samples.}\label{table1}
    \label{tab:hogehoge}
    \begin{tabular}{|c|c|}
        \hline
         &$\left(\overline{\sum_{\text{space}}S}\right)_{\text{first}} \ /\ \left(\overline{\sum_{\text{space}}S}\right)_{\text{second}}$\\ \hline
        $N=8$ & 0.41659733$\pm$0.03954734\\ \hline
        $N=10$ & 0.42963458$\pm$0.02621359\\ \hline
        $N=12$ & 0.43605110$\pm$0.02234253 \\ \hline
    \end{tabular}
\end{table}
\normalsize
From this result, we assume that the following relationship holds:
\begin{equation}
\label{relation1}
\left(\overline{\sum_{\text{space}}S}\right)_{\text{first}}:\left(\overline{\sum_{\text{space}}S}\right)_{\text{second}}\sim\quad   1:2
\end{equation}
The theoretical basis of Eq. (\ref{relation1}) is discussed in Appendix \ref{Ap2}.
\normalsize
Even when performing truncation that retains only $\chi$ singular values, the same calculation can be performed as in the case without truncation.
First, we calculate the maximum value of the sum of EE in the space direction when truncated at $\chi$, denoted as $\left(\sum_{\text{space}}S^{\chi} \right)_{\text{max}}$. This corresponds to a bond dimension distribution where the bond dimension increases from the edges toward the center in the MPS as $2,4,8,\cdots, \chi, \cdots, \chi$, reaching $\chi$ at some point and remaining at $\chi$ thereafter. 
Following the same approach as Eq. (\ref{Smax theory}), the maximum value of the sum of EE in the space direction is given as follows:

\footnotesize
\begin{equation}
\begin{split}
\label{Smax theory2}
    &\left(\sum_{\text{space}}S^{\chi} \right)_{\text{max}}\\
    =&2\sum_{k=1}^{\lfloor\log_2{\chi}\rfloor} \log_2{2^k}+2\left(\frac{N}{2}-\lfloor\log_2{\chi}\rfloor-1\right)\log_2{\chi}+\log_2{\chi}\\
    =&\lfloor\log_2{\chi}\rfloor(\lfloor\log_2{\chi}\rfloor+1)+\log_2{\chi}(N-2\lfloor\log_2{\chi}\rfloor-1)
\end{split}
\end{equation}
\normalsize

\begin{figure}[h]
  \centering\vspace{(0mm)}
  \includegraphics[width=0.85\linewidth]{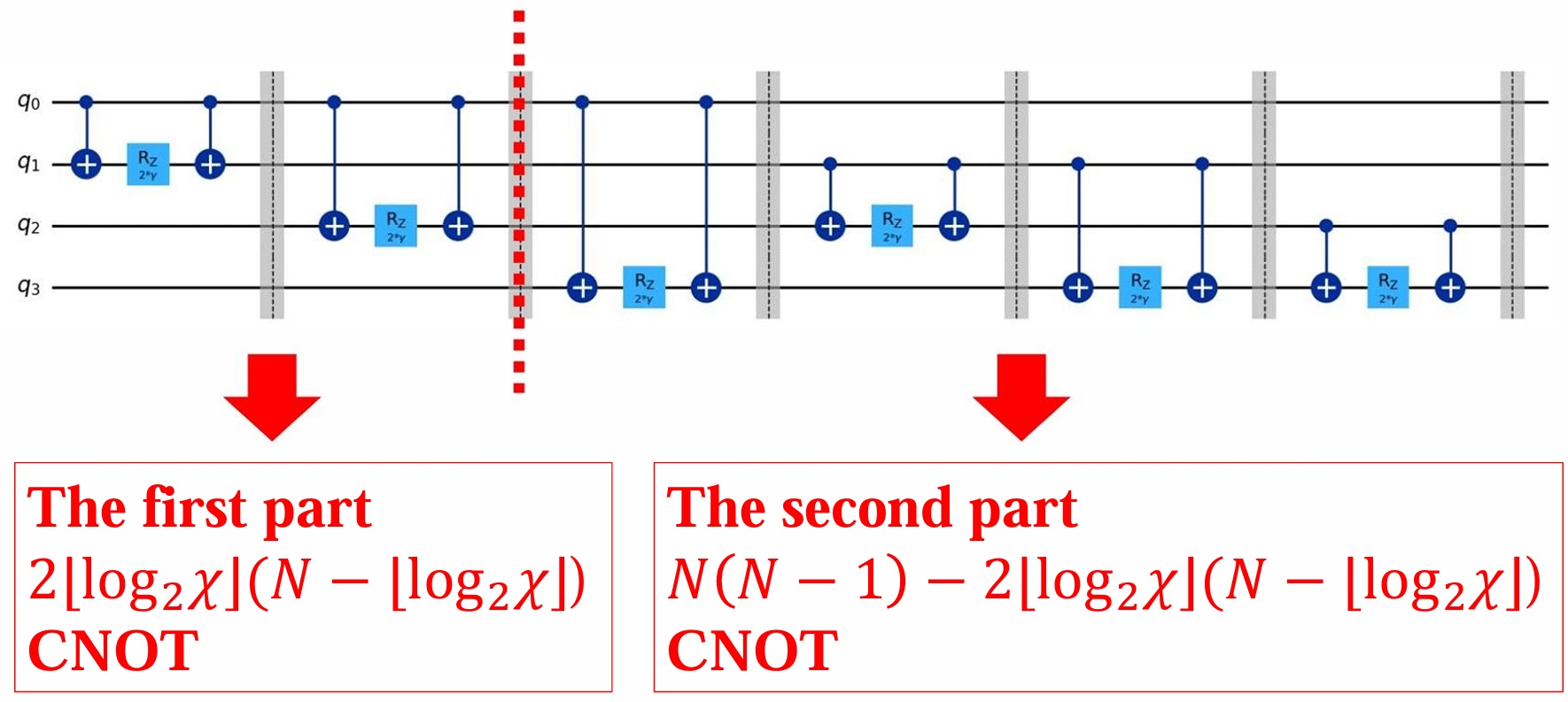}
  \caption{An illustrative diagram of how to divide the QAOA circuit into the first and second parts in the case with truncation.}
  \label{fig:first and second 2}
\end{figure}

Consider dividing the circuit into two parts with respect to the number of CNOT gates in the circuit, as in the case without truncation. However, in the case of truncation, instead of splitting the circuit exactly in half, the first $2\lfloor\log_2{\chi}\rfloor(N-\lfloor\log_2{\chi}\rfloor)$ CNOT gates and the remaining CNOT gates are used to divide the circuit into two parts (refer to Fig. \ref{fig:first and second 2}). This number represents the required number of CNOT gates for all bond dimensions to reach their maximum value in the case where the bond dimension increases at the shortest possible rate.
Then, for each of these divisions, calculate the average value of $\sum_{\text{space}}S^{\chi}$. These averages are denoted as $(\overline{\sum_{\text{space}}S^{\chi}})_{\text{first}}$ and $(\overline{\sum_{\text{space}}S^{\chi}})_{\text{second}}$ for the first and second parts, respectively.

From the numerical calculation results with truncation for 
$N=8,10,12$, it was found that a ratio similar to the one in Table \ref{table1} is obtained. Table \ref{table2} shows the results of these numerical calculations.

\Large
\begin{table}[h]
    \centering
    \caption{Verification in numerical calculations. The average is taken over 100 samples.}\label{table2}
    \label{tab:hogehoge}
    \begin{tabular}{|c|c|}
        \hline
         &$\left(\overline{\sum_{\text{space}}S^{\chi}}\right)_{\text{first}} \ /\ \left(\overline{\sum_{\text{space}}S^{\chi}}\right)_{\text{second}}$\\ \hline
        $N=8$  ,\ $\chi$=4 & 0.51182160$\pm$0.07255609\\ \hline
        $N=8$  ,\ $\chi$=8 & 0.45979109$\pm$0.04391587\\ \hline
        $N=10$  ,\ $\chi$=2 & 0.55000148$\pm$0.11751299\\ \hline
        $N=10$  ,\ $\chi$=16 & 0.46330448$\pm$0.02809180\\ \hline
        $N=12$ ,\ $\chi$=8 & 0.43605110$\pm$0.02234253\\ \hline
    \end{tabular}
\end{table}
\normalsize
From this result, we assume that the following relationship holds:
\begin{equation}
\label{relation2}
\left(\overline{\sum_{\text{space}}S^{\chi}}\right)_{\text{first}}:\left(\overline{\sum_{\text{space}}S^{\chi}}\right)_{\text{second}}\sim\quad   1:2
\end{equation}
The theoretical basis of Eq. (\ref{relation2}) is also discussed in Appendix \ref{Ap2}.

Furthermore, it was found that using the numerical calculation results for both the cases with and without truncation, results similar to those in the Table \ref{table3} are obtained.
This result shows that the ratio between the average of the sum of EE in the second part and the theoretical maximum value of the sum of EE remains the same regardless of whether truncation is applied or not.

\begin{table}[h]
    \centering
    \caption{Verification in numerical calculations. The average is taken over 100 samples.}\label{table3}
    \scalebox{1}{
    \begin{tabular}{|c|c|}
        \hline
         \rule{0pt}{1.5em} &$\qquad  \frac{\left(\overline{\sum_{\text{space}}S^{\chi}}\right)_{\text{second}}/\left(\sum_{\text{space}}S^{\chi} \right)_{\text{max}} }{ \left(\overline{\sum_{\text{space}}S}\right)_{\text{second}}  /\left(\sum_{\text{space}}S \right)_{\text{max}}} \qquad $\\ \hline
        $N=8$  ,\ $\chi$=4 & 0.94552621$\pm$0.08692903\\ \hline
        $N=8$  ,\ $\chi$=8  & 1.03091806$\pm$0.02700337\\ \hline
        $N=10$  ,\ $\chi$=2 & 0.84383451$\pm$0.14901796\\ \hline
        $N=10$  ,\ $\chi$=16 & 1.02427734$\pm$0.01551792\\ \hline
        $N=12$  ,\ $\chi$=8 & 0.92339795$\pm$0.10971339\\ \hline
    \end{tabular}
    }
\end{table}

\normalsize
From this result, we assume that the following relationship holds:
\begin{equation}
\label{become 1}
\frac{\left(\overline{\sum_{\text{space}}S^{\chi}}\right)_{\text{second}}/\left(\sum_{\text{space}}S^{\chi} \right)_{\text{max}} }{ \left(\overline{\sum_{\text{space}}S}\right)_{\text{second}}  /\left(\sum_{\text{space}}S \right)_{\text{max}}}\quad \sim \quad 1 
\end{equation}
Rearranging Eq. (\ref{become 1}) results in the following:
\begin{equation}
\label{relation3}
\frac{\left(\overline{\sum_{\text{space}}S}\right)_{\text{second}}}{\left(\sum_{\text{space}}S \right)_{\text{max}}} \ \sim \ 
\frac{\left(\overline{\sum_{\text{space}}S^{\chi}}\right)_{\text{second}} }{ \left(\sum_{\text{space}}S^{\chi} \right)_{\text{max}}  }\ \equiv \ \alpha
\end{equation}
The theoretical basis of Eq. (\ref{relation3}) is also discussed in Appendix \ref{Ap2}.

Using the previous discussion, an analytical calculation of Eq. (\ref{Scaling for EE}) becomes possible.
By combining Eqs. (\ref{relation1}), (\ref{relation2}), and (\ref{relation3}), it can be confirmed that the following two relational expressions hold.
\small
\begin{equation}
\label{ratio1}
\left(\overline{\sum_{\text{space}}S}\right)_{\text{first}}:\left(\overline{\sum_{\text{space}}S}\right)_{\text{second}}:\left(\sum_{\text{space}}S \right)_{\text{max}}\sim\  \frac{\alpha}{2}:\alpha:1
\end{equation}

\begin{equation}
\label{ratio2}
\left(\overline{\sum_{\text{space}}S^{\chi}}\right)_{\text{first}}:\left(\overline{\sum_{\text{space}}S^{\chi}}\right)_{\text{second}}:\left(\sum_{\text{space}}S^{\chi}\right)_{\text{max}}\sim\  \frac{\alpha}{2}:\alpha:1
\end{equation}
\normalsize

First, consider the case without truncation.
Using Eqs. (\ref{ratio1}) and (\ref{Smax theory}), the time direction summation of $\sum_{\text{space}} S$ over the entire circuit can be approximately computed (denoted as $\sum_{\text{space\ \&\ time}} S$). Approximating $\frac{N(N-1)}{2} \sim \frac{N^2}{2}$, we compute $\sum_{\text{space\ \&\ time}} S$ as follows:
\footnotesize
\begin{equation}
\label{S without}
\begin{split}
        \sum_{\text{space\ \&\ time}}S &\sim \frac{N^2}{2} \left(\overline{\sum_{\text{space}}S}\right)_{\text{first}} +\frac{N^2}{2} \left(\overline{\sum_{\text{space}}S}\right)_{\text{second}} \\
        &\sim \frac{\alpha}{2}\cdot \frac{N^2}{2} \left(\sum_{\text{space}}S \right)_{\text{max}} + \alpha \cdot \frac{N^2}{2} \left(\sum_{\text{space}}S \right)_{\text{max}} \\
        &=\frac{3}{16}\alpha N^4
\end{split}
\end{equation}
\normalsize

Next, consider the case with truncation.
Using Eq. (\ref{Smax theory2}) and assuming that Eq. (\ref{ratio2}) holds approximately, the time direction summation of $\sum_{\text{space}} S^{\chi}$ can be computed approximately in the same way as Eq. (\ref{S without}). The calculation is as follows (where we use $N(N-1) \sim N^2$):
\footnotesize
\begin{equation}
\label{S with}
        \sum_{\text{space\ \&\ time}}S^{\chi}
        \sim  \alpha\left(\sum_{\text{space}}S^{\chi} \right)_{\text{max}}\left( N^2-N\lfloor\log_2{\chi}\rfloor+(\lfloor\log_2{\chi}\rfloor)^2    \right)
\end{equation}
\normalsize

From Eqs. (\ref{S without}) and (\ref{S with}), the theoretical approximation of Eq. (\ref{Scaling for EE}) can be obtained as follows:

\begin{widetext}
\begin{equation}
\label{result1}
\begin{split}
    \frac{\sum S^{\chi}}{\sum S}&\sim\frac{\sum_{\text{space\ \&\ time}}S^{\chi}}{\sum_{\text{space\ \&\ time}}S}\\
    &=\frac{16}{3}\cdot \frac{\left\{\lfloor\log_2{\chi}\rfloor(\lfloor\log_2{\chi}\rfloor+1)+\log_2{\chi}(N-2\lfloor\log_2{\chi}\rfloor-1)\right\}\left\{ N^2-N\lfloor\log_2{\chi}\rfloor+(\lfloor\log_2{\chi}\rfloor)^2    \right\}}{N^4}\\
&\sim \ \frac{16}{3}\cdot\frac{\log_2{\chi}}{N} \left(1-\frac{\log_2{\chi}}{N} \right)\left(1-\frac{\log_2{\chi}}{N} +\left( \frac{\log_2{\chi}}{N}\right)^2 \right)\\
        &\sim \ \mathcal{H}\left(\frac{2\log_2{\chi}}{N} \right)
\end{split}
\end{equation}
\end{widetext}
where the approximation of $\lfloor\log_2{\chi}\rfloor\sim \log_2{\chi}$ was used in the third line. 

\begin{figure}[h]
  \centering\vspace{(0mm)}
  \includegraphics[width=0.9\linewidth]{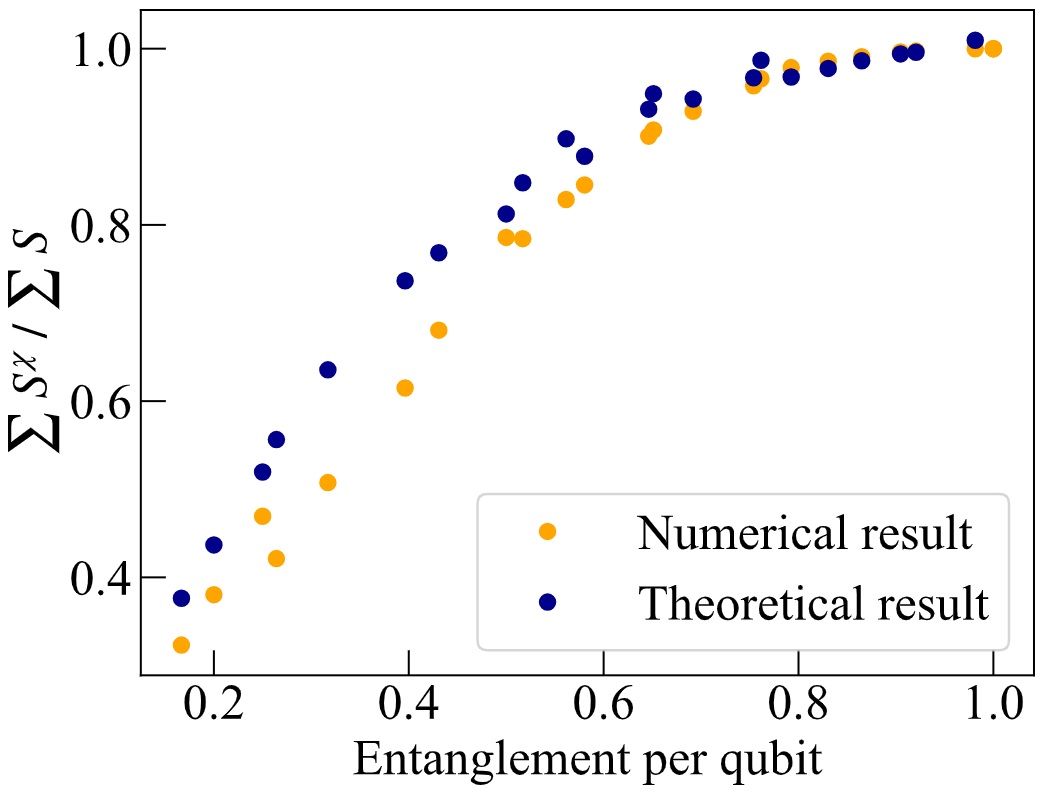}
  \caption{The scaling relations of Eq. (\ref{Scaling for EE}) (numerical calculation, orange) and  Eq. (\ref{result1}) (theoretical approximation, darkblue) are overlaid (an example with one QAOA layer in a complete graph).}
  \label{fig:compare}
\end{figure}

Finally, we present a plot of the numerical calculation results of Eq. (\ref{Scaling for EE}), along with the theoretical approximation given by the second line of Eq. (\ref{result1}) (Fig. \ref{fig:compare}).
The two curves exhibit generally similar trends, but a noticeable discrepancy can
be observed between them. Possible reasons for the discrepancy include: (1) the errors in the relationships in Tables \ref{table1}, \ref{table2} and \ref{table3} were too large, (2) the approximations of $\frac{N(N-1)}{2} \sim \frac{N^2}{2}$ in Eq. (\ref{S without}) and $N(N-1) \sim N^2$ in Eq. (\ref{S with}) were too rough, and (3) errors in the numerical calculations, among others.

However, it is important that a theoretical analysis of the scaling relations has been provided for the first time. By correcting the incomplete aspects of this theory, it is likely possible to obtain theoretical values that match the numerical calculation results. The connection between the scaling relation presented in this study and the one shown in \cite{Scaling1} is unclear, but it will likely help clarify the theoretical foundation of the scaling relation in \cite{Scaling1}.

\section{conclusion and outlook}
In this study, we developed the scaling relations presented in previous research \cite{Scaling1} and discovered that a new scaling relation based on the entanglement entropy (EE) holds. Moreover, since this scaling relation is based on the clearly defined quantity of entanglement entropy, theoretical analysis is possible. In this study, we propose a new function for the scaling relation.
Additionally, from the results of numerical calculations used to derive the scaling function, several interesting relationships regarding the entanglement entropy in the QAOA circuit were obtained. This phenomenon is likely to hold specifically after optimizing the parameters in the QAOA circuit. Moreover, if other quantum algorithms exhibit similar behavior of entanglement entropy, there is a possibility that scaling relations will hold, similar to QAOA. It would be interesting to investigate this. Also, regarding the relationship with previous studies, the question of why the ratio of the energy expectation value and the sum of the EE leads to a similar scaling relation remains unresolved, and its investigation will be a task for future research.

\section*{ACKNOWLEDGMENTS}
We are grateful for the support from JST Moonshot R$\&$D (JPMJMS2061). We would like to thank T. Kadowaki and Y. Suzuki for their valuable advice at the National Institute of Advanced Industrial Science and Technology (AIST). We acknowledge the use of IBM Quantum services for this work. A part of this work was performed for Council for Science, Technology and Innovation (CSTI), Cross-
ministerial Strategic Innovation Promotion Program (SIP), “Promoting the application of advanced
quantum technology platforms to social issues”(Funding agency : QST).

\appendix
\renewcommand{\appendixname}{APPENDIX }

\section{Details of the simulation}
\label{Simulation1}
Here, we describe the details of the simulations performed. First, for the circuit simulation and optimization of QAOA, we used the AerSimulator in IBM's Qiskit \cite{qiskit}. Furthermore, for the simulation of quantum circuits in MPS, we utilized QUIMB \cite{quimb}. The initial parameters for QAOA were randomly chosen from the range 
$[-\pi, \pi]$. The optimization method used was 'COBYLA', with a tolerance of 
\textit{tol}=1\textit{e}-13. All data presented in this paper were obtained by sampling 100 instances of random edge weights ($w_{i,j}\in [0,\ 1]$) in a complete graph and calculating their average.

\section{Reason for taking the summation}
\label{Ap1}
In the following, we discuss whether the scaling relation holds when considering the ratio of EE at a single bond without summing over EE.
The performed simulation involves numerically calculating the EE at the central bond (the bond at the $N/2$ position from the edge, where $N$ is the number of qubits) in the MPS immediately after applying the final CNOT gate in the QAOA circuit. We then take the ratio of EE with and without truncation.
This quantity differs from the entanglement entropy scaling relation presented in \cite{Scaling1}, where the average entanglement entropy at the central bond is divided by $N$.
As a result, we confirm that the scaling relation does not hold. Expressing this mathematically, we obtain the following equation ($\lambda_{i}$ represents the singular values):
\begin{equation}
\label{eq:EE for single point}
    \frac{S^{\chi}}{S}=\frac{-\sum^{\chi}_{i=1} \lambda^2_{i}\log_2{\lambda^2_{i}}}{-\sum^{2^{N/2}}_{i=1} \lambda^2_{i}\log_2{\lambda^2_{i}}}\ne \mathcal{G}\left(\frac{2\log_2{\chi}}{N}\right)
\end{equation}

\begin{figure}[h]
  \centering\vspace{(0mm)}
  \includegraphics[width=0.9\linewidth]{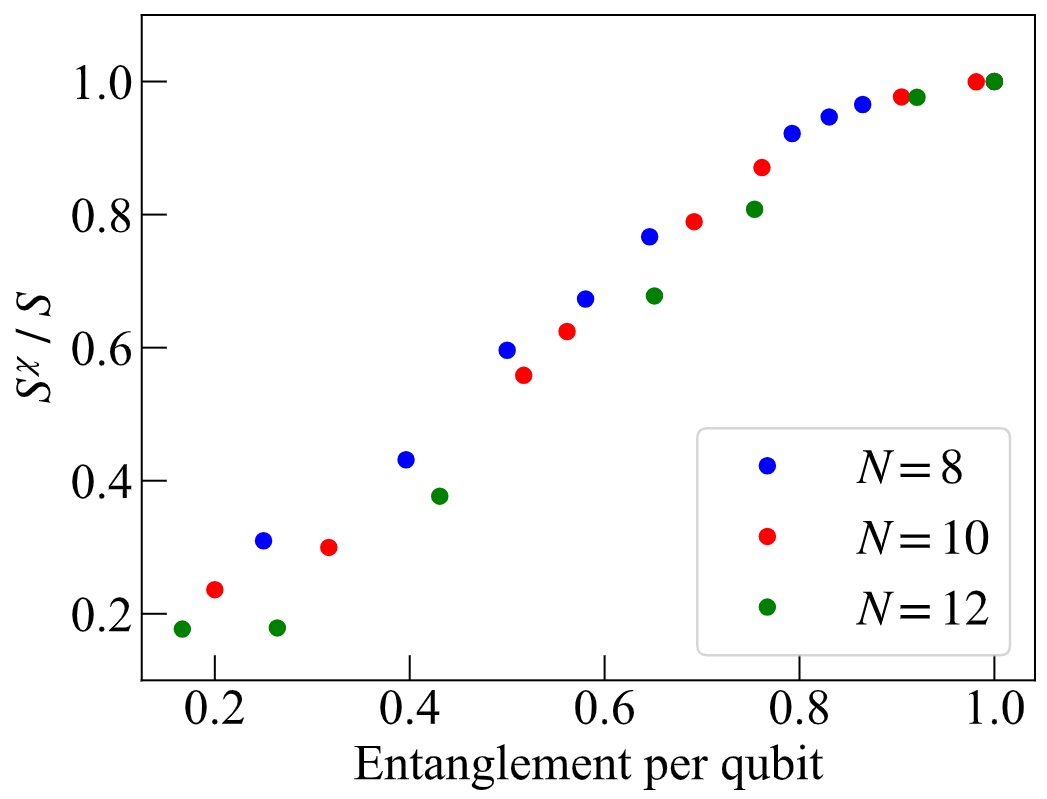}
  \caption{Violation of the scaling relation in the case of Eq. (\ref{eq:EE for single point}) (an example with a complete graph and one QAOA layer). The average is taken over 100 samples.}
  \label{fig:EE for single point}
\end{figure}

The graph corresponding to Eq. (\ref{eq:EE for single point}) is Fig. \ref{fig:EE for single point}.

\begin{figure*}[t]
  \centering\vspace{(0mm)}
  \includegraphics[width=\linewidth]{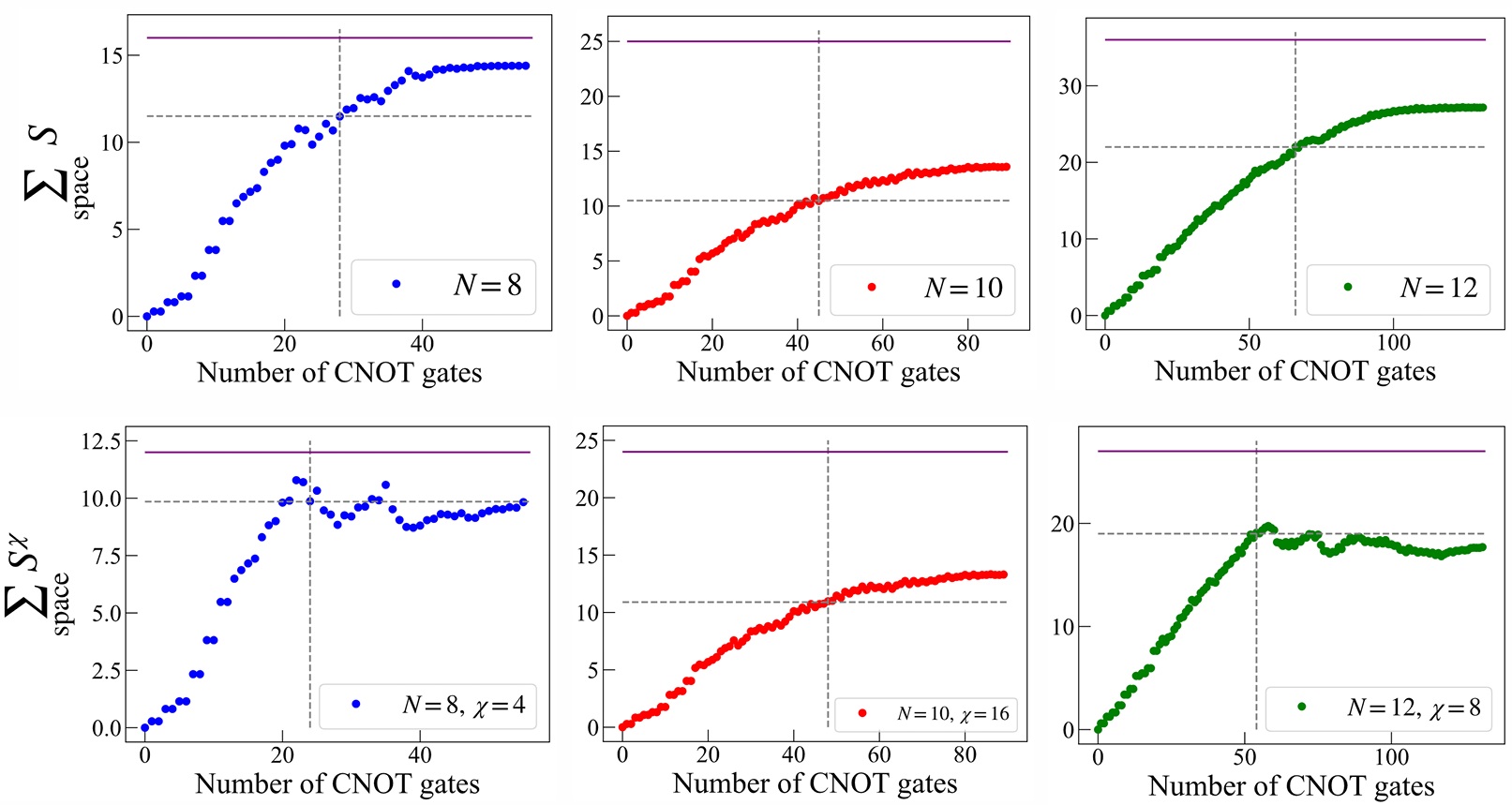}
  \caption{Time evolution of the sum of entanglement in the space direction in the QAOA circuit (an example with a complete graph and one QAOA layer). The average is taken over 100 samples. The top row represents the case without truncation, while the bottom row represents the case with truncation.
The purple straight line indicates the theoretical maximum value in each graph, calculated based on Eqs. (\ref{Smax theory}) and (\ref{Smax theory2}).
The gray vertical line in the top row marks the point where the number of CNOT gates in the QAOA circuit reaches half, while in the bottom row, it indicates the point where the number of CNOT gates reaches $2\lfloor\log_2{\chi}\rfloor(N-\lfloor\log_2{\chi}\rfloor)$. The gray horizontal line is drawn for visual clarity.}
  \label{fig:EE for quantum circuit}
\end{figure*}

Observing Fig. \ref{fig:EE for single point}, we see that when the horizontal axis $\frac{2\log_2{\chi}}{N}$ is close to 1, the data collapse onto a single curve independent of $N$. However, as the horizontal axis approaches 0, $N$-dependence emerges, and the scaling relation breaks down (for smaller $N$, $S_{\chi}/S$ becomes larger). Therefore, this provides an example where the scaling relation does not hold without summation.

A fundamental reason for taking the summation is related to the transformation process from the original complete graph (two-dimensional) to the MPS representation (one-dimensional). Since the complete graph is forcibly converted into an MPS representation, looking at only a single bond fails to capture the correct overall entanglement properties.
Additionally, summation can be considered equivalent to averaging (although averaging typically involves division by the total number, this factor cancels out when taking ratios). For these reasons, focusing on the total sum (or average) of EE across the entire circuit, rather than EE at a single bond, is a reasonable approach.

\section{Analyzing the relationship in Entanglement Entropy}
\label{Ap2}
The theoretical basis of Eqs. (\ref{relation1}) and (\ref{relation2}) is presumed to be related to the development of EE within the circuit. As seen in Fig. \ref{fig:EE for quantum circuit}, the sum of EE in the space direction increases linearly at the beginning of the QAOA circuit, regardless of $N$, and gradually approaches a constant value when the number of CNOT gates exceeds half (top row) or $2\lfloor\log_2{\chi}\rfloor(N-\lfloor\log_2{\chi}\rfloor)$ (bottom row).
From this fact, Eqs. (\ref{relation1}) and (\ref{relation2}) can be explained. 

The results in Fig. \ref{fig:EE for quantum circuit} partially align with the discussion in \cite{Scaling2}. The fact that the EE in Fig. \ref{fig:EE for quantum circuit} does not decrease suggests that QAOA is not approaching the exact solution due to the limited number of QAOA layers. That is, the scaling relations in Eq. (\ref{relation1}) and (\ref{relation2}) are likely to hold only when the solution is not close to the exact one, and the scaling relation in Eq. (\ref{Scaling for EE}) is valid under these conditions. In \cite{Scaling2}, it was reported that the scaling relations hold only when the number of QAOA layers is small. The connection between the scaling relation in this study and the one in \cite{Scaling2} is unclear, but it is interesting that there are similarities.

The theoretical foundation of Eq. (\ref{relation3}) is explained as follows.
If truncation is performed, the amount of EE is limited. As a result, $\left(\overline{\sum_{\text{space}}S^{\chi}}\right)_{\text{second}}$ and $\left(\sum_{\text{space}}S^{\chi} \right)_{\text{max}}$ become smaller in proportion to $\chi$. Consequently, when taking the ratio, it can be interpreted that the result remains the same as in the case without truncation.

In summary, the universality concerning $N$ in the scaling relation can be interpreted as arising from the universality of EE growth (i.e., EE follows a similar developmental pattern in all cases, independent of $N$).


\begin{thebibliography}{9}
\bibitem{Feynman}
R. P. Feynman, Simulating physics with computers, in \textit{Feynman and Computation} (CRC Press, Boca Raton, FL, 2018), pp. 133-153.

\bibitem{QC1}
S. K. Sood, and Pooja, Quantum Computing Review: A Decade of Research, \href{https://ieeexplore.ieee.org/document/10167529}{IEEE Transactions on Engineering Management (Volume: 71) (2023).} 

\bibitem{QC2}
M. A. Shafique, A. Munir, I. Latif, Quantum Computing: Circuits, Algorithms, and Applications, \href{https://ieeexplore.ieee.org/document/10423012}{IEEE Access (Volume: 12). (2024).} 


\bibitem{QC3}
A. W. Harrowand and A. Montanaro, Quantum computational supremacy, \href{https://www.nature.com/articles/nature23458}{Nature 549, 203 (2017)} 

\bibitem{IBMQ}
IBM Quantum (2025), \href{https://quantum.ibm.com/services/resources}{https://quantum.ibm.com/} 

\bibitem{Error correction}
D. Gottesman, The Heisenberg Representation of Quantum Computers, \href{https://arxiv.org/abs/quant-ph/9807006}{arXiv:quant-ph/9807006.}

\bibitem{EC2}
J. Roffe, Quantum error correction: an introductory guide, \href{https://www.tandfonline.com/doi/full/10.1080/00107514.2019.1667078}{Taylor $\&$ Francis, Contemporary Physics 2019, VOL. 60, NO. 3, 226–245 (2019).} 

\bibitem{EC3}
Google Quantum AI and Collaborators, Quantum error correction below the surface code threshold, \href{https://www.nature.com/articles/s41586-024-08449-y}{Nature, volume 638, p. 920–926 (2025).} 

\bibitem{NISQ}
J. Preskill, Quantum Computing in the NISQ era and beyond, \href{https://quantum-journal.org/papers/q-2018-08-06-79/}{Quantum 2, 79 (2018).}

\bibitem{Grover}
L. K. Grover, A fast quantum mechanical algorithm for database search, Proceedings, \href{https://dl.acm.org/doi/10.1145/237814.237866}{28th Annual ACM Symposium on the Theory of Computing, p. 212 (1996).}

\bibitem{NC}
M. A. Nielsen and I. L. Chuang, \textit{Quantum Computation and Quantum Information 10th Anniversary Edition}, Cambridge University Press, Cambridge, 2012.

\bibitem{Shor}
P. W. Shor, Polynomial-Time Algorithms for Prime Factorization and Discrete Logarithms on a Quantum Computer, \href{https://epubs.siam.org/doi/10.1137/S0097539795293172}{SIAM Journal on Computing. 26 , 1484 (1997).}


\bibitem{TN}
K. Okunishi, T. Nishino and H. Ueda, Developments in the Tensor Network — from Statistical Mechanics to Quantum Entanglement, \href{https://journals.jps.jp/doi/10.7566/JPSJ.91.062001}{Phys. Soc. Jpn. 91, 062001 (2022).}

\bibitem{TN2}
T. L. Patti, J. Kossaifi, A. Anandkumar, and S. F. Yelin, Variational quantum optimization with multibasis encodings, \href{https://journals.aps.org/prresearch/abstract/10.1103/PhysRevResearch.4.033142}{Physical Review Research 4, 033142 (2022).} 

\bibitem{TN3}
B. A. Martin, T. Ayral, F. Jamet, M. J. Rančić, and P. Simon, Combining matrix product states and noisy quantum computers for quantum simulation, \href{https://journals.aps.org/pra/abstract/10.1103/PhysRevA.109.062437}{PRA 109, 062437 (2024).} 


\bibitem{SVD}
J. Bisgard, \textit{Analysis and Linear Algebra: The Singular Value Decomposition and Applications. Student Mathematical Library}, AMS. ISBN 978-1-4704-6332-8 (2021).

\bibitem{QAOA1}
E. Farhi, J. Goldstone, S. Gutmann, A Quantum Approximate Optimization Algorithm, \href{https://arxiv.org/abs/1411.4028}{arXiv:1411.4028.}

\bibitem{QAOA2}
L. Zhou, S. Wang, S. Choi, H. Pichler and M. D. Lukin, Quantum Approximate Optimization Algorithm, Performance, Mechanism, and Implementation on Near-Term Devices, 
\href{https://journals.aps.org/prx/abstract/10.1103/PhysRevX.10.021067}{PRX 10, 021067 (2020).}

\bibitem{Scaling1}
M. Dupont , N. Didier, M. J. Hodson, J. E. Moore, M. J. Reagor, Calibrating the Classical Hardness of the Quantum Approximate Optimization Algorithm, \href{https://journals.aps.org/prxquantum/abstract/10.1103/PRXQuantum.3.040339}{PRX QUANTUM 3, 040339 (2022).}

\bibitem{MPS}
D. Perez-Garcia, F. Verstraete, M.M. Wolf, J.I. Cirac, Matrix Product State Representations, \href{https://arxiv.org/abs/quant-ph/0608197}{arXiv:0608197v2.}


\bibitem{Scaling2}
A. C. Nakhl, T. Quella, M. Usman, Calibrating the role of entanglement in variational quantum circuits,  \href{https://journals.aps.org/pra/abstract/10.1103/PhysRevA.109.032413}{PRA 109,032413 (2024).}

\bibitem{qiskit}
H. Abraham et al., Qiskit: An Open-source Framework for Quantum Computing, (2019), \href{https://github.com/Qiskit}{https://github.com/Qiskit}

\bibitem{quimb}
J. Gray, quimb: A python package for quantum information and many-body calculations, \href{https://joss.theoj.org/papers/10.21105/joss.00819}{J. Open Source Softw. 3, 819 (2018)}
\end{thebibliography}
\end{document}